\documentclass[10pt]{article}
\usepackage{cite}
\usepackage{epsfig}
\usepackage[dvips,usenames]{color}
\usepackage{color}
\usepackage{rotating}
\usepackage{amsmath,amssymb,amsfonts}
\usepackage{latexsym}
\begin{document}
\setcounter{section}{0}
\setcounter{equation}{0}
\setcounter{figure}{0}
\setcounter{table}{0}
\setcounter{footnote}{0}
\begin{flushright}
CP3-09-36
\end{flushright}

\vspace{10pt}

\begin{center}
{\bf\Large The Klauder--Daubechies Construction}

\vspace{5pt}

{\bf\Large of the Phase Space Path Integral and}

\vspace{5pt}

{\bf\Large the Harmonic Oscillator}
\end{center}
\vspace{10pt}
\begin{center}
Jan Govaerts$^{\dagger,\ddagger,}$\footnote{Fellow of the Stellenbosch Institute
for Advanced Study (STIAS), 7600 Stellenbosch, Republic of South Africa.},
Calvin Matondo Bwayi$^\star$
and Olivier Mattelaer$^{\dagger,\diamond}$\\
\vspace{15pt}
$^\dagger${\sl Center for Particle Physics and Phenomenology (CP3),\\
Institut de Physique Nucl\'eaire, Universit\'e catholique de Louvain (U.C.L.),\\
2, Chemin du Cyclotron, B-1348 Louvain-la-Neuve, Belgium}\\
{\it E-Mail: Jan.Govaerts@uclouvain.be, Olivier.Mattelaer@uclouvain.be}\\
\vspace{10pt}
$^\ddagger${\sl International Chair in Mathematical Physics and Applications (ICMPA-UNESCO Chair),\\
University of Abomey--Calavi, 072 B. P. 50, Cotonou, Republic of Benin}\\
\vspace{10pt}
$^\star${\sl Department of Physics, University of Kinshasa (UNIKIN),\\
Kinshasa, Democratic Republic of Congo}\\
{\it E-mail: matymatondo@yahoo.fr}\\
\vspace{10pt}
$^\diamond${\sl Istituto Nazionale di Fisica Nucleare (INFN), Sezione di Roma Tre,\\
and Dipartimento di Fisica``Edoardo Amaldi", Universit\`a degli Studi Roma Tre, I-00146, Roma, Italy}
\end{center}

\vspace{5pt}

\begin{quote}
The canonical operator quantisation formulation corresponding to the Klauder--Daubechies construction of the
phase space path integral is considered. This formulation is explicitly applied and solved in the case of the harmonic
oscillator, thereby illustrating in a manner complementary to Klauder and Daubechies' original work some of the promising features
offered by their construction of a quantum dynamics. The Klauder--Daubechies functional integral involves a regularisation
parameter eventually taken to vanish, which defines a new physical time scale. When extrapolated to the field
theory context, besides providing a new regularisation of short distance divergences, keeping a finite value for
that time scale offers some tantalising prospects when it comes to strong gravitational quantum systems.
\end{quote}

\vspace{5pt}

\setcounter{footnote}{0}

\section{Introduction}
\label{KD1.Sec1}

The central r\^ole of the phase space symplectic one-form in the quantisation programme is well
known and understood. No less important and crucial to the physical properties of the quantum system however,
is the r\^ole of an implicit phase space Riemannian metric---albeit
a ``shadow phase space metric"\cite{KD1.Klauder1,KD1.Klauder2,KD1.Klauder2.2,KD1.Klauder3,KD1.Klauder4,KD1.Klauder5}
for what the classical world is concerned. This is convincingly argued by John R. Klauder
in an insightful and thought provoking paper\cite{KD1.Klauder1} deserving to be much more widely
known, which relies on prior work with Ingrid Daubechies\cite{KD1.KD1,KD1.KD2,KD1.KD3,KD1.KD4,KD1.KD5,KD1.KD6}.
Making the r\^ole of this phase space Riemannian metric explicit circumvents the ambiguities and difficulties
inherent to the formal path integral definition of a quantised system. It even provides for a perfectly well defined functional
integral over continuous paths in phase space. One requirement these two geometrical structures on phase space have to meet is that
the symplectic and Riemannian metrics both define an identically normalised phase space volume form.

The Klauder--Daubechies construction is achieved through stochastic calculus methods,
involving a Wiener measure of which the diffusion parameter provides a regulator which, when eventually taken to infinity,
reproduces the correct quantum mechanical amplitudes obeying the Schr\"odinger equation of the system. In terms closer
to a physicist's intuition perhaps, this Wiener measure is associated to the statistical Brownian motion of a particle
propagating in the background Riemannian geometry of phase space, with a specific diffusion time scale taken eventually
to vanish. One of the remarkable features of the Klauder--Daubechies construction of the phase space path integral is
its inherent manifest covariance under general canonical transformations in phase space, much in
contradistinction to all other approaches leading to path integral representations of quantum amplitudes, or even
to naive canonical operator quantisation rules.

In order to make these statements somewhat more explicit, for the purpose of this introductory discussion only let us assume
a system of physical units such that all relevant parameters and scales are set to unity, inclusive of $\hbar=1$,
and let us restrict to a single degree of freedom system, $q(t)\in\mathbb{R}$, with canonically conjugate momentum,
$p(t)\in\mathbb{R}$, obeying at the quantum level the Heisenberg algebra,
$\left[Q,P\right]=i\mathbb{I}$, $Q^\dagger=Q$, $P^\dagger=P$.
If $|q,p\rangle$ denote the normalised canonical Weyl--Heisenberg quantum coherent states
labelled by all classical phase space states and associated to the normalised Fock vacuum $|0\rangle$ such that
$\left(Q+iP\right)|0\rangle=0$ \cite{KD1.KS}, the Klauder--Daubechies path integral (KD-PI) representation of
the associated matrix elements of the quantum system's evolution operator of Hamiltonian $\hat{H}_0(Q,P)$ is given
in the form\cite{KD1.Klauder1},
\begin{eqnarray}
\langle q_f,p_f|e^{-i T\hat{H}_0}|q_i,p_i\rangle &=& \lim_{\tau_0\rightarrow 0^+}
e^{C_0\frac{T}{\tau_0}}\int_{(q(t_i),p(t_i))=(q_i,p_i)}^{(q(t_f),p(t_f))=(q_f,p_f)}
\left[\frac{{\cal D}q(t){\cal D}p(t)}{2\pi}\right]\times \nonumber \\
&&\qquad \times e^{i\int_{t_i}^{t_f}dt\left[\frac{1}{2}\left(\dot{q}p-q\dot{p}\right)-h(q,p)\right]}
\times e^{-\tau_0\int_{t_i}^{t_f}dt\frac{1}{2}\left(\dot{q}^2+\dot{p}^2\right)},
\label{eq:KDPI1}
\end{eqnarray}
where the time interval, $T=t_f-t_i$, is such that $T>0$, and $\tau_0>0$ is a time scale regularisation parameter.
In this expression one integrates over all those paths in phase space possessing as end points those classical states
associated to the external quantum coherent states. Furthermore, $h(q,p)$ is the coherent state symbol representing
the Hamiltonian operator through\cite{KD1.KS}
\begin{equation}
\hat{H}_0=\int_{(\infty)}\frac{dq\,dp}{2\pi}|q,p\rangle\,h(q,p)\,\langle q,p|,\qquad
\int_{(\infty)}\frac{dq\,dp}{2\pi}|q,p\rangle\,\langle q,p|=\mathbb{I}.
\end{equation}
To lowest order in $\hbar$, $h(q,p)$ coincides with the classical Hamiltonian, $H_0(q,p)$. One thus recognizes in the phase
of the first exponential factor inside the path integral the first-order Hamiltonian action of the system, inclusive
of quantum corrections in $h(q,p)$. In the absence of these corrections, that phase factor provides the usual naive
formal definition of the quantised system through the phase space path integral, a definition however, which is not
free of ambiguities nor difficulties, among which a lack of covariance under phase space canonical transformations.
Note in particular the contribution to that Hamiltonian action in $dt(\dot{q} p-q \dot{p})/2=(dq\, p - q\, dp)/2=K$,
which defines the symplectic one-form of the phase space symplectic geometry. The associated volume form, $\omega=dK=dp\wedge dq$,
is thus normalised to unity.

However, the expression in (\ref{eq:KDPI1}) carries still two further $\tau_0$ dependent exponential factors.
Returning to the very first one later on, the very last one inside the path integral is of a purely statistical character,
being purely real gaussian, in contradistinction to the previous pure phase factor of a purely quantum mechanical character.
In effect the real gaussian factor plays the r\^ole of a phase space Wiener measure which regularises, for any finite $\tau_0>0$,
the ordinary naive path integral based on the purely imaginary (gaussian and higher order) phase factor alone.
Furthermore one recognizes in that real gaussian contribution precisely the Brownian motion of a particle in the background phase space euclidean
geometry associated to the Weyl--Heisenberg algebra defined by the operators $Q$ and $P$ (other
homogeneous geometries are also discussed in Ref.\cite{KD1.Klauder1}).
Note that the volume element associated to that Riemannian geometry with as metric a tensor given by the unit matrix,
is again normalised to unity, as is the volume form associated to the symplectic one-form involved in the pure phase factor.

Note well that by having introduced the time scale $\tau_0$, the one dimensional system with configuration space
coordinate $q$ and two dimensional phase space $(q,p)$ has been promoted to some two dimensional system with
configuration space $(q,p)$, hence a four dimensional phase space, of which the dimensional reduction back to the space
$q$ only is achieved through the limit $\tau_0\rightarrow 0^+$. As such the Lagrangian action for this effective two dimensional system reads,
\begin{equation}
\int_{t_i}^{t_f}dt\left[\frac{1}{2}i\tau_0\left(\dot{q}^2+\dot{p}^2\right)+\frac{1}{2}\left(\dot{q}p-q\dot{p}\right)
-h(q,p)\right].
\label{eq:action1}
\end{equation}
In this expression one recognizes the action of a particle of pure positive imaginary mass, $m_0=i\tau_0$, moving
in a two dimensional euclidean plane, subjected to a potential energy $h(q,p)$, as well as a velocity dependent,
hence magnetic, coupling defined by the symplectic
one-form of the Hamiltonian formulation of the original system, as if the particle were coupled to a static homogeneous
magnetic field perpendicular to the two dimensional space $(q,p)$. Except for the mass factor which is not real, this is precisely
a generalised Landau problem in phase space with interaction energy $h(q,p)$ \cite{KD1.Klauder2,KD1.Klauder2.2}. As is well known,
in the absence of this interaction energy, the energy levels of the quantised Landau problem are organised in
infinitely degenerate discrete Landau levels, with a gap set by the ratio of the magnetic coupling to the mass.
In the presence of the interaction energy $h(q,p)$, the Landau level degeneracies are lifted but states are still organised in
discrete Landau sectors with a gap set by the same ratio. In the limit of a vanishing mass, namely in the present context
the limit $\tau_0\rightarrow 0^+$, this gap grows infinite. Hence in order that not all Landau sectors decouple
one has to adjust the quantum vacuum energy of the lowest Landau sector such that the energy of all states in
that sector retain a finite energy in that limit. This is precisely the reason for the very first exponential factor
in (\ref{eq:KDPI1}) multiplying the path integral, $C_0$ being some normalisation factor to be adjusted 
accordingly (which may be done up to an arbitrary finite contribution even when $h(q,p)=0$ \cite{KD1.Klauder1}).
And as a consequence, the surviving quantum
states of the lowest Landau sector span the quantum Hilbert space of the original quantum system with the single degree of freedom $q$.
Dimensional reduction in phase space is achieved for the quantum system by projecting onto its lowest Landau sector the extended
quantised system, as defined by (\ref{eq:KDPI1}). Note that by the same token noncommutativity in the $(q,p)$ space
is induced once again through that projection, out of commuting operators $(q,p)$ as configuration space coordinates for the
extended dynamics. In essence, this is the genesis of the noncommutative Moyal plane of noncommutative quantum mechanics as well.

Incidentally, besides this intriguing possibility of having ``extra dimensions" introduced in a dynamics
which are neither of a space- nor a time-like character (as in Kaluza-Klein or string theory contexts) but  are rather
of a phase space character, the mixture of both purely quantum and statistical behaviours present in the
formulation of a quantum dynamics as provided by the KD-PI construction in (\ref{eq:KDPI1}), reminds one of
progress made by G. 't Hooft\cite{KD1.tHooft} with precisely such motivations in mind towards a deterministic formulation
of quantum dynamics displaying at the same time a stochastic behaviour.

Until recently\cite{KD1.GovMatt,KD1.Matt} to the present authors' best knowledge, and in spite of all the potential
interest offered by this approach to quantum dynamics, if only to illustrate explicitly the workings of (\ref{eq:KDPI1})
no actual evaluation of the KD-PI was available---certainly not for a finite value for $\tau_0$---even for as simple
a test-bed system as the harmonic oscillator, the basis for all of perturbative relativistic quantum field theory.
Certainly to the authors of the KD-PI is it clear---having proved it---that the quantum dynamics of the original system is
recovered in the limit $\tau_0=0$. But if the formulation is to find practical applications, some
explicit evaluations with finite $\tau_0$ are most presumably useful. Furthermore, besides the path integral point of view
on which the construction of (\ref{eq:KDPI1}) is based, a complementary understanding of the quantum properties
of the extended system associated to the action in (\ref{eq:action1}) from the canonical operator quantisation
point of view should prove to be of relevance as well, and could lead to further insight into the workings
of the limit $\tau_0=0$. Finally, keeping the value for $\tau_0$ finite may also be of interest in the context of
deformations of algebraic structures associated to quantum dynamics in a more general setting,
for instance that of quantum gravity and noncommutative geometries of spacetime\cite{KD1.GovMatt,KD1.Matt,KD1.SG}.

The purpose of the present paper is not to justify the result in (\ref{eq:KDPI1}), but rather, by starting from it,
to show explicitly that it indeed reproduces the correct quantum dynamics of the harmonic oscillator,
and thereby acquire greater familiarity with the meaning of the Klauder--Daubechies
approach and the prospects it may offer. And since this has already been done in Refs.\cite{KD1.GovMatt,KD1.Matt}
through a direct saddle point evaluation of the path integral (\ref{eq:KDPI1}) for a finite $\tau_0$ and in the limit $\tau_0=0$,
the same issue is addressed here directly from the canonical
operator quantisation point of view, based on the $\tau_0$ deformed effective action (\ref{eq:action1}) of the system defined
over the original phase space promoted to a two dimensional configuration space. In the case of the one dimensional
harmonic oscillator, one is thus dealing with a Landau problem with pure positive imaginary mass subjected to
a harmonic potential well. Even though the quantum solution for that system should be straightforward enough,
its lack of unitarity and its properties under the limit $\tau_0=0$ are sufficiently instructive to deserve
a detailed analysis. At the same time, a broader and perhaps clearer understanding of the relevance and potential interest of
the Klauder--Daubechies construction of the phase space path integral is achieved.

The paper is organised as follows. In Section 2, the canonical formulation associated to the extended action (\ref{eq:action1})
is constructed. Section 3 then applies this formalism to the one dimensional harmonic oscillator to construct the
canonical quantisation of its extended formulation, and its quantum solution, enabling thereby an explicit analysis of the
limit $\tau_0=0$ corresponding to the effective projection onto the lowest Landau sector of the system. Section 4 then
addresses the evaluation of the projected quantum evolution operator for a finite value of $\tau_0$, to compare
with the saddle point evaluation of Refs.\cite{KD1.GovMatt,KD1.Matt}. Finally, some conclusions are presented in Section 5.

\section{Canonical Formulation of the Extended System}
\label{KD1.Sec2}

First let us still consider an arbitrary one degree of freedom system, with canonically conjugate
phase space variables $(q,p)$ and classical Hamiltonian $H_0(q,p)$, and reinstate dimensionful quantities,
inclusive of all explicit factors of $\hbar$.
In order to account for the different physical dimensions of the configuration space variable, $q$, and its
conjugate momentum, $p$, let us also introduce a constant factor $\lambda_0$ having the dimension of mass times
angular frequency. It proves then useful to work in terms of the following rescaled phase space coordinates,
$\phi^a$ ($a=1,2$), with
\begin{equation}
\phi^1=\frac{1}{\sqrt{\lambda_0}}p,\qquad
\phi^2=q\sqrt{\lambda_0},
\end{equation}
having the canonical Poisson brackets, $\left\{\phi^a,\phi^b\right\}=-\epsilon^{ab}$, $\epsilon^{ab}$ being the
two dimensional antisymmetric symbol with $\epsilon^{12}=+1$. Hence the classical Hamiltonian first-order action
of the system reads,
\begin{equation}
S_0[\phi^a]=\int dt\left[\frac{1}{2}\left(\dot{q}p - q\dot{p}\right)-H_0(q,p)\right]=
\int dt\left[\frac{1}{2}\epsilon_{ab}\phi^a\dot{\phi}^b-H_0(\phi^a)\right].
\end{equation}

For what the extended system is concerned, the function $H_0(\phi^a)$ gets replaced by the
symbol $h(\phi^a)=h(q,p)$, while the associated Lagrangian action reads,
\begin{equation}
S[\phi^a]=\int dt\left[\frac{1}{2}i\tau_0\delta_{ab}\dot{\phi}^a\dot{\phi}^b+\frac{1}{2}\epsilon_{ab}\phi^a\dot{\phi}^b-h(\phi^a)+E_0\right],
\end{equation}
$\delta_{ab}$ being the phase space euclidean metric, and $E_0$ some ($\hbar$ dependent) constant to be adjusted later on
in order to retain quantum states of finite energy in the limit $\tau_0\rightarrow 0^+$ in the manner explained previously.

Developing a classical canonical formulation corresponding to this Lagrangian action as such is problematic.
Indeed, even when initial or boundary conditions for $\phi^a$ are specified to be real valued, because of the pure imaginary mass
term trajectories solving the associated classical Euler--Lagrange equations of motions are bound to become complex valued,
hence also the momentum variables, $p_a$, conjugate to the configuration space ones, $\phi^a$. At the quantum level it would
therefore appear to be unjustified to associate to both these quantities operators that are self-adjoint.

However, one should keep in mind that the above Lagrangian action for the extended system only contributes
inside a quantum path integral of the form,
\begin{equation}
\int\left[{\cal D}\phi^a(t)\right]\,e^{\frac{i}{\hbar}S[\phi^a]},
\end{equation}
where integration is taken over real paths in the real valued configuration space, $\phi^a(t)$, not involving therefore
the complex valued classical trajectories (unless one considers an evaluation of the integral through contour deformations
into the complex plane, as is done effectively in a saddle point evaluation \cite{KD1.GovMatt,KD1.Matt}).
In terms of this path integral it becomes possible to introduce
real valued variables, $p_a$, canonically conjugate to the real configuration space ones, $\phi^a$, as auxiliary
variables for some well defined real gaussian integrals, thereby bringing the path integral into canonical first-order form,
namely,
\begin{equation}
\int\left[{\cal D}\phi^a(t)\right]\,e^{\frac{i}{\hbar}S[\phi^a]}=
\int\left[{\cal D}\phi^a(t){\cal D}p_a(t)\right]e^{\frac{i}{\hbar}\int dt\left[\dot{\phi}^a p_a - H(\phi^a,p_a)\right]},
\end{equation}
where
\begin{equation}
H(\phi^a,p_a)=\frac{1}{2i\tau_0}\delta^{ab}\left(p_a+\frac{1}{2}\epsilon_{ac}\phi^c\right)
\left(p_b+\frac{1}{2}\epsilon_{bd}\phi^d\right)+h(\phi^a)-E_0
\end{equation}
(the absolute normalisation of the functional integration measures is left unspecified).
In particular note that the gaussian integrals over $p_a$ are real and well defined precisely because the
mass parameter, $m_0=i\tau_0$, is pure positive imaginary.

Clearly it is this latter form of the path integral which defines the canonical formulation of the extended system,
with as real canonically conjugate phase space variables $(\phi^a,p_a)$ and canonical Hamiltonian the function $H(\phi^a,p_a)$.
As is well known, such a path integral is associated to an operator realisation over some Hilbert space providing
a representation of the following extended Heisenberg algebra,
\begin{equation}
\left[\hat{\phi}^a,\hat{p}_b\right]=i\hbar\delta^a_b\mathbb{I},\qquad
\hat{\phi}^{a\dagger}=\hat{\phi}^a,\qquad \hat{p}_a^\dagger=\hat{p}_a,
\label{eq:extalg}
\end{equation}
with indeed hermitian operators, and note well, also commuting $\hat{\phi}^a$, namely $\hat{q}$ and $\hat{p}$ operators.
Hence rather than consider the path integral in (\ref{eq:KDPI1}),
an equivalent realisation of the same extended quantum system for a finite $\tau_0$ value is defined
by this operator algebra and the quantum Hamiltonian
\begin{equation}
\hat{H}=\frac{1}{2i\tau_0}\delta^{ab}\left(\hat{p}_a+\frac{1}{2}\epsilon_{ac}\hat{\phi}^c\right)
\left(\hat{p}_b+\frac{1}{2}\epsilon_{bd}\hat{\phi}^d\right)+h(\hat{\phi}^a)-E_0.
\end{equation}
It is thus the eigenspectrum of this operator that needs to be understood as a function of $\tau_0$,
as well as its behaviour in the limit $\tau_0=0$. Note well however that because of the pure imaginary mass
parameter, this operator is not hermitian, $\hat{H}^\dagger\ne\hat{H}$, hence the quantum dynamics of the extended quantum system is
not unitary, for any finite $\tau_0>0$. In particular, its eigenspectrum proves to be complex but
with a dependence on $\tau_0$ such that for those states that survive the limit $\tau_0=0$, their limiting
energy eigenvalues are real once again and coincide with the eigenspectrum of the original unitary quantised system.
This very point may thus be studied explicitly for a function $h(\phi^a)$ which, for example, is purely quadratic
(and linear) in the variables $\phi^a$, namely essentially the case of the one dimensional harmonic oscillator.

Let us henceforth consider a harmonic oscillator of mass $m$ and angular frequency $\omega_0>0$, with
then the choice $\lambda_0=m\omega_0$. The Hamiltonian then reads
\begin{equation}
H_0(q,p)=\frac{1}{2m}p^2+\frac{1}{2}m\omega^2_0 q^2=\frac{1}{2}\omega_0\delta_{ab}\phi^a \phi^b.
\end{equation}
Except for an additive constant proportional to $\hbar$ which may be absorbed in the choice for $E_0$,
in this case the symbol $h(q,p)$ for the quantum Hamiltonian $\hat{H}_0$ coincides with $H_0(q,p)$.
Consequently, the operator quantisation of the extended system in the case of the harmonic oscillator
is defined by the Heisenberg algebra in (\ref{eq:extalg}) as well as the following quantum Hamiltonian,
\begin{equation}
\hat{H}=\frac{1}{2i\tau_0}\delta^{ab}\left(\hat{p}_a+\frac{1}{2}\epsilon_{ac}\hat{\phi}^c\right)
\left(\hat{p}_b+\frac{1}{2}\epsilon_{bd}\hat{\phi}^d\right)+\frac{1}{2}\omega_0\delta_{ab}\hat{\phi}^a\hat{\phi}^b-E_0,
\label{eq:HHO}
\end{equation}
the diagonalisation of which we now address.

\section{The Ordinary Harmonic Oscillator}
\label{KD1.Sec3}

\subsection{A bi-module of Fock-like algebras}

Given the hermitian operators $\hat{\phi}^a$ and $\hat{p}_a$, let us introduce first the following Fock operators,
\begin{equation}
a_a=\frac{1}{2\sqrt{\hbar}}\left(\hat{\phi}^a+2i\hat{p}_a\right),\qquad
a^\dagger_a=\frac{1}{2\sqrt{\hbar}}\left(\hat{\phi}^a-2i\hat{p}_a\right),
\end{equation}
which define the tensor product of two Fock algebras,
\begin{equation}
\left[a_a, a^\dagger_b\right]=\delta_{ab}\mathbb{I}.
\end{equation}
Next, consider the following helicity Fock operators,
\begin{equation}
a_\pm=\frac{1}{\sqrt{2}}\left(a_1 \mp i a_2\right),\qquad
a^\dagger_\pm=\frac{1}{\sqrt{2}}\left(a^\dagger_1 \pm i a^\dagger_2\right),
\end{equation}
such that,
\begin{equation}
\left[a_\pm, a^\dagger_\pm\right]=\mathbb{I},\qquad
\left[a_\pm, a^\dagger_\mp\right]=0.
\end{equation}
The inverse relations are,
\begin{eqnarray}
\hat{\phi}^1=\sqrt{\frac{\hbar}{2}}\left(a_+ + a_- + a^\dagger_+ + a^\dagger_- \right) &,&
\hat{p}_1=-\frac{i}{2}\sqrt{\frac{\hbar}{2}}\left(a_+ + a_- - a^\dagger_+ - a^\dagger_- \right), \nonumber \\
\hat{\phi}^2=i\sqrt{\frac{\hbar}{2}}\left(a_+ - a_- - a^\dagger_+ + a^\dagger_- \right) &,&
\hat{p}_2=\frac{1}{2}\sqrt{\frac{\hbar}{2}}\left(a_+ - a_- + a^\dagger_+ - a^\dagger_- \right),
\end{eqnarray}
with in particular,
\begin{equation}
\hat{p}_1+\frac{1}{2}\hat{\phi}^2=-i\sqrt{\frac{\hbar}{2}}\left(a_- - a^\dagger_- \right),\qquad
\hat{p}_2-\frac{1}{2}\hat{\phi}^1=-\sqrt{\frac{\hbar}{2}}\left(a_- + a^\dagger_- \right).
\end{equation}

To construct an abstract representation of these algebraic structures, consider now a normalised Fock vacuum, $|\Omega\rangle$,
for the helicity Fock operators,
\begin{equation}
a_\pm |\Omega\rangle=0,\qquad \langle\Omega|\Omega\rangle=1,
\end{equation}
with the following orthonormalised states spanning the Hilbert space of the extended quantum system,
\begin{equation}
|n_+,n_-;\Omega\rangle=\frac{1}{\sqrt{n_+!\, n_-!}}\left(a^\dagger_+\right)^{n_+}
\left(a^\dagger_-\right)^{n_-}|\Omega\rangle,\qquad
\langle n_+,n_-;\Omega|m_+,m_-;\Omega\rangle=\delta_{n_+,m_+}\,\delta_{n_-,m_-},
\end{equation}
where $n_+,n_-=0,1,2,\ldots$, hence with the resolution of the unit operator,
\begin{equation}
\sum_{n_+,n_-=0}^\infty |n_+,n_-;\Omega\rangle \langle n_+,n_-;\Omega|=\mathbb{I}.
\end{equation}

Incidentally these are the states that diagonalise the quantum operator $\hat{H}$ in the absence of the
interaction coupling $h(\hat{\phi}^a)$, leading to Landau levels labelled by $n_-=0,1,\ldots$ and
infinitely degenerate in $n_+=0,1,\ldots$ In particular, the lowest Landau level, $|n_+,n_-=0;\Omega\rangle$
($n_+=0,1,\ldots$), will turn out to define the subspace of the Hilbert space of the extended quantum system which coincides
with the Hilbert space of the original quantum system, namely the lowest Landau sector in presence
of the interaction energy $h(\hat{\phi}^a)$ in the limit when $\tau_0\rightarrow 0^+$.
For that reason it is useful to already introduce the projector
onto that subspace of quantum states of the extended system,
\begin{equation}
\mathbb{P}_0=\sum_{n_+=0}^\infty |n_+,n_-=0;\Omega\rangle\langle n_+,n_-=0;\Omega|,\qquad
\mathbb{P}_0^2=\mathbb{P}_0,\qquad \mathbb{P}^\dagger_0=\mathbb{P}_0.
\end{equation}
Note that we then have for the projected operators generating the Heisenberg algebra in the extended Hilbert space,
\begin{equation}
\mathbb{P}_0\left(\hat{p}_1+\frac{1}{2}\hat{\phi}^2\right)\mathbb{P}_0=0,\qquad
\mathbb{P}_0\left(\hat{p}_2-\frac{1}{2}\hat{\phi}^1\right)\mathbb{P}_0=0,
\end{equation}
showing that after projection only the projected coordinates $\mathbb{P}_0\hat{\phi}^a\mathbb{P}_0=\bar{\phi}^a$
are independent operators, with as commutation relations,
\begin{equation}
\left[\bar{\phi}^a,\bar{\phi}^b\right]=-i\hbar\epsilon^{ab}\mathbb{P}_0.
\end{equation}
Hence indeed on the projected subspace one recovers the Heisenberg algebra of the original quantum system,
even though within the extended Hilbert space the phase space position operators $\hat{\phi}^a$ commute with each other.

However the above states do not diagonalise the total Hamiltonian $\hat{H}$ in presence of the interaction $h(\hat{\phi}^a)$,
even for the harmonic oscillator. In the latter case, other linear combinations of the basic
operators $\hat{\phi}^a$ and $\hat{p}_a$ are required.  For that purpose, let us introduce two specific real quantities, $R_0$
and a phase $\varphi_0$, defined by the following relation,
\begin{equation}
R^2_0\,e^{2i\varphi_0}\equiv 1+4i\omega_0\tau_0,\qquad R_0>0,\qquad 0\le\varphi_0<\frac{\pi}{4},
\label{eq:R0P0}
\end{equation}
as well as the complex variable, $\rho$, and its complex conjugate, $\bar{\rho}$,
\begin{equation}
\rho=\sqrt{R_0}\,e^{\frac{1}{2}i\varphi_0},\qquad \bar{\rho}=\sqrt{R_0}\,e^{-\frac{1}{2}i\varphi_0}.
\end{equation}
Note that in the limit $\tau_0\rightarrow 0^+$, or in the absence of the coupling $\omega_0$, $R_0$ and $\rho$
both go to unity while $\varphi_0$ vanishes.

Consider then the operators,
\begin{equation}
A_a=\frac{1}{2\sqrt{\hbar}}\left(\rho\hat{\phi}^a+\frac{2i}{\rho}\hat{p}_a\right),\qquad
B_a=\frac{1}{2\sqrt{\hbar}}\left(\rho\hat{\phi}^a-\frac{2i}{\rho}\hat{p}_a\right),
\end{equation}
as well as their adjoints,
\begin{equation}
A^\dagger_a=\frac{1}{2\sqrt{\hbar}}\left(\bar{\rho}\hat{\phi}^a-\frac{2i}{\bar{\rho}}\hat{p}_a\right),\qquad
B^\dagger_a=\frac{1}{2\sqrt{\hbar}}\left(\bar{\rho}\hat{\phi}^a+\frac{2i}{\bar{\rho}}\hat{p}_a\right),
\end{equation}
which are such that,
\begin{equation}
\left[A_a,B_b\right]=\delta_{ab}\mathbb{I}=\left[B^\dagger_a,A^\dagger_b\right].
\end{equation}
Had it not been for the fact that $\rho$ is a complex quantity, the operators $A_a$ and $B_a$ would
have been adjoints of one another. We have, for instance,
\begin{equation}
A^\dagger_a=\cos\varphi_0\,B_a\,-\,i\sin\varphi_0\,A_a,\qquad
B^\dagger_a=\cos\varphi_0\,A_a\,-\,i\sin\varphi_0\,B_a.
\end{equation}
Furthermore the operators $A_a$ and $B_a$ almost coincide with $a_a$ and $a^\dagger_a$
above, respectively, but only if $\rho=1$, namely whenever $\omega_0\tau_0=0$. Clearly, the operators
$A_a$, $B_a$ and their adjoints may be expressed as linear combinations of $a_a$ and $a^\dagger_a$.

Finally let us introduce the helicity combinations,
\begin{eqnarray}
A_\pm=\frac{1}{\sqrt{2}}\left(A_1\mp i A_2\right) &,&
B_\pm=\frac{1}{\sqrt{2}}\left(B_1 \pm i B_2\right), \nonumber \\
A^\dagger_\pm=\frac{1}{\sqrt{2}}\left(A^\dagger_1 \pm i A^\dagger_2\right) &,&
B^\dagger_\pm=\frac{1}{\sqrt{2}}\left(B^\dagger_1 \mp i B^\dagger_2\right),
\end{eqnarray}
such that
\begin{equation}
\left[A_\pm, B_\pm\right]=\mathbb{I}=\left[B^\dagger_\pm, A^\dagger_\pm\right],
\label{eq:Focklike}
\end{equation}
as well as,
\begin{eqnarray}
A^\dagger_\pm=\cos\varphi_0\,B_\pm\,-\,i\sin\varphi_0\,A_\mp &,&
A_\pm=\cos\varphi_0\,B^\dagger_\pm\,+\,i\sin\varphi_0\,A^\dagger_\mp, \nonumber \\
B^\dagger_\pm=\cos\varphi_0\,A_\pm\,-\,i\sin\varphi_0\,B_\mp &,&
B_\pm=\cos\varphi_0\,A^\dagger_\pm\,+\,i\sin\varphi_0\,B^\dagger_\mp.
\end{eqnarray}
Expressing these operators in terms of $a_\pm$ and $a^\dagger_\pm$, one finds,
\begin{eqnarray}
A_\pm=\frac{\rho+\rho^{-1}}{2}\,a_\pm\,+\,\frac{\rho-\rho^{-1}}{2}\,a^\dagger_\mp &,&
A^\dagger_\pm=\frac{\bar{\rho}+\bar{\rho}^{-1}}{2}\,a^\dagger_\pm\,+\,\frac{\bar{\rho}-\bar{\rho}^{-1}}{2}\,a_\mp, \nonumber \\
B_\pm=\frac{\rho+\rho^{-1}}{2}\,a^\dagger_\pm\,+\,\frac{\rho-\rho^{-1}}{2}\,a_\mp &,&
B^\dagger_\pm=\frac{\bar{\rho}+\bar{\rho}^{-1}}{2}\,a_\pm\,+\,\frac{\bar{\rho}-\bar{\rho}^{-1}}{2}\,a^\dagger_\mp.
\label{eq:relations}
\end{eqnarray}

The Fock-like algebraic relations in (\ref{eq:Focklike}) are very much similar to those of ordinary Fock algebras, except
for the fact that the operators $B_\pm$ and $A_\pm$ (on the one hand, or their adjoints on the other hand) are not adjoints
of one another. Yet,
a representation theory may be constructed in very much the same way, leading to dual states we shall refer to
as $A$- and $B$-Fock states. This representation is built on $A$- and $B$-Fock vacua, $|\Omega_A\rangle$ and $|\Omega_B\rangle$,
respectively, such that
\begin{equation}
A_\pm|\Omega_A\rangle=0,\qquad
B^\dagger_\pm|\Omega_B\rangle=0.
\end{equation}
By an appropriate choice of phases and normalisations, it is always possible to assume that the
inner product of these two states is set to unity,
\begin{equation}
\langle\Omega_A|\Omega_B\rangle=1=\langle\Omega_B|\Omega_A\rangle.
\end{equation}
The $A$-Fock states are then defined by
\begin{equation}
|N_+,N_-;\Omega_A\rangle=\frac{1}{\sqrt{N_+!\,N_-!}}B^{N_+}_+ B^{N_-}_-|\Omega_A\rangle,
\end{equation}
while for the $B$-Fock states,
\begin{equation}
|N_+,N_-;\Omega_B\rangle=\frac{1}{\sqrt{N_+!\,N_-!}}\left(A^\dagger_+\right)^{N_+}\left(A^\dagger_-\right)^{N_-}
|\Omega_B\rangle,
\end{equation}
where $N_+,N_-=0,1,2,\ldots$ As a matter of fact, since the operators $A_\pm$, $B_\pm$ and their adjoints are
linear combinations of the Fock operators $a_\pm$ and $a^\dagger_\pm$, it is clear that either set of states,
$|N_+,N_-;\Omega_A\rangle$ or $|N_+,N_-;\Omega_B\rangle$, spans the entire Hilbert space of the quantum extended
system. More specifically, each of these two sets provides a basis of that space, these two bases being in fact
dual to one another,
\begin{equation}
\langle N_+,N_-;\Omega_A|M_+,M_-;\Omega_B\rangle=\delta_{N_+,M_+}\delta_{N_-,M_-}=
\langle N_+,N_-;\Omega_B|M_+,M_-;\Omega_A\rangle.
\end{equation}
Consequently, one also has the following resolutions of the unit operator,
\begin{equation}
\sum_{N_+,N_-=0}^\infty|N_+,N_-;\Omega_A\rangle\langle N_+,N_-;\Omega_B|=\mathbb{I}=
\sum_{N_+,N_-=0}^\infty|N_+,N_-;\Omega_B\rangle\langle N_+,N_-;\Omega_A|.
\end{equation}
In other words, the three sets of states, $|n_+,n_-;\Omega\rangle$, $|N_+,N_-;\Omega_A\rangle$ and
$|N_+,N_-;\Omega_B\rangle$, define three different bases of the same extended Hilbert space, with the
basis $|n_+,n_-;\Omega\rangle$ being self-dual since orthonormalised, while the other two bases are dual to one another.

Note that the action of the $A_\pm$ and $B_\pm$ operators on the $A$-Fock states, on the one hand, and of
the $B^\dagger_\pm$ and $A^\dagger_\pm$ operators on the $B$-Fock states, on the other hand, is precisely like that of
ordinary annihilation and creation Fock operators, respectively, on ordinary Fock states. In particular,
the $A$-Fock states $|N_+,N_-;\Omega_A\rangle$ (resp., $B$-Fock states $|N_+,N_-;\Omega_B\rangle$) are
eigenstates of the operators $B_\pm A_\pm$ (resp., $A^\dagger_\pm B^\dagger_\pm$) with eigenvalues $N_\pm$.

Given the identities (\ref{eq:relations}) relating the different Fock-like operators, it should be clear that
the relations between these three different bases are obtained as Bogoliubov transformations. Introducing the
complex parameter
\begin{equation}
\lambda=\frac{\rho - \rho^{-1}}{\rho + \rho^{-1}},\qquad
\bar{\lambda}=\frac{\bar{\rho}-\bar{\rho}^{-1}}{\bar{\rho}+\bar{\rho}^{-1}},
\end{equation}
a little analysis shows that the $A$- and $B$-Fock vacua are given as,
\begin{equation}
|\Omega_A\rangle=\left(\frac{2}{\rho+\rho^{-1}}\right)\,e^{-\lambda a^\dagger_+ a^\dagger_-}\,|\Omega\rangle,\qquad
|\Omega_B\rangle=\left(\frac{2}{\bar{\rho}+\bar{\rho}^{-1}}\right)\,
e^{-\bar{\lambda}a^\dagger_+ a^\dagger_-}\,|\Omega\rangle,
\end{equation}
and similarly,
\begin{equation}
|\Omega_B\rangle=N_B(\varphi_0)\,e^{i\tan\varphi_0\,B_+ B_-}\,|\Omega_A\rangle,\qquad
|\Omega_A\rangle=N_A(\varphi_0)\,e^{-i\tan\varphi_0\,A^\dagger_+ A^\dagger_-}\,|\Omega_B\rangle,
\end{equation}
$N_A(\varphi_0)$ and $N_B(\varphi_0)$ being two normalisation factors whose evaluation is not required here,
\begin{equation}
N^{-1}_A(\varphi_0)=\langle\Omega_B|e^{-i\tan\varphi_0\,A^\dagger_+ A^\dagger_-}|\Omega_B\rangle,\qquad
N^{-1}_B(\varphi_0)=\langle\Omega_A|e^{i\tan\varphi_0\,B_+ B_-}|\Omega_A\rangle.
\end{equation}
These different representations relating the different Fock vacua as coherent helicity pairing excitations of one another,
thus establish that indeed all three sets of Fock states provide complete bases of the same extended Hilbert space
in which to diagonalise the total quantum Hamiltonian $\hat{H}$.

Finally, note that in the limit where $\tau_0\rightarrow 0^+$, all three sets of Fock states then coalesce into
a single set, namely the states $|n_+,n_-;\Omega\rangle$ ($n_+,n_-=0,1,\ldots$), since then all three Fock vacua
become identical to $|\Omega\rangle$ while we have the following correspondences for the creation and annihilation
operators,
\begin{equation}
A_\pm\rightarrow a_\pm,\qquad
B_\pm\rightarrow a^\dagger_\pm,\qquad
A^\dagger_\pm\rightarrow a^\dagger_\pm,\qquad
B^\dagger_\pm\rightarrow a_\pm.
\end{equation}

\subsection{The energy spectrum}

With the previous representation theory of the extended Hilbert space at hand, diagonalisation of the total
Hamiltonian (\ref{eq:HHO}) of the extended system is readily achieved. In terms of the operators introduced above,
a little substitution easily finds,
\begin{equation}
\hat{H}=\hbar\frac{R_0e^{i\varphi_0}+1}{2i\tau_0}\,B_- A_-\,+\,\hbar\frac{R_0e^{i\varphi_0}-1}{2i\tau_0}\,
\left(B_+ A_+ + \frac{1}{2}\right)\,+\,
\left(\hbar\frac{R_0e^{i\varphi_0}+1}{4i\tau_0}-E_0\right).
\end{equation}
Obviously, the $A$-Fock states, $|N_+,N_-;\Omega_A\rangle$, are the eigenstates of that operator, while those
of its adjoint, $\hat{H}^\dagger\ne\hat{H}$, are the $B$-Fock states, $|N_+,N_-;\Omega_B\rangle$. Furthermore
the subtraction constant $E_0$ needs to be adjusted as follows,
\begin{equation}
E_0=\hbar\frac{R_0e^{i\varphi_0}+1}{4i\tau_0}\,-\,\Delta E_0(\omega_0,\tau_0),\qquad
\lim_{\tau_0\rightarrow 0^+} \Delta E_0(\omega_0,\tau_0)=0,
\end{equation}
where the function $\Delta E_0(\omega_0,\tau_0)$ is {\it a priori} otherwise arbitrary (it may even be complex for a
finite value of $\tau_0$), and in fact is of the form,
\begin{equation}
\Delta E_0(\omega_0,\tau_0)=\hbar\omega_0\Delta{\cal E}_0(\omega_0\tau_0),
\end{equation}
$\Delta{\cal E}_0(\omega_0\tau_0)$ being a function only of the product $(\omega_0\tau_0)$ which vanishes
when that argument vanishes. In the limit $\tau_0\rightarrow 0^+$, clearly then
only the lowest Landau sector with $N_-=0$ retains finite energy values, namely the states
$|N_+,N_-=0;\Omega_A\rangle\rightarrow |n_+=N_+,n_-=0;\Omega\rangle$ for $\hat{H}$ and
$|N_+,N_-=0;\Omega_B\rangle\rightarrow |n_+=N_+,n_-=0;\Omega\rangle$ for $\hat{H}^\dagger$, with
$N_+=0,1,\ldots$

Given that choice for the subtraction constant $E_0$, the complex energy spectrum of the system, for a finite
value of $\tau_0>0$, is given as,
\begin{equation}
\hat{H}|N_+,N_-;\Omega_A\rangle= E(N_+,N_-)|N_+,N_-;\Omega_A\rangle,\qquad
\hat{H}^\dagger|N_+,N_-;\Omega_B\rangle=\bar{E}(N_+,N_-)|N_+,N_-;\Omega_B\rangle,
\end{equation}
with
\begin{equation}
E(N_+,N_-)=\hbar\frac{R_0e^{i\varphi_0}+1}{2i\tau_0}\,N_-\,+\,
\hbar\frac{R_0e^{i\varphi_0}-1}{2i\tau_0}\,\left(N_+ +\frac{1}{2}\right)+\Delta E_0(\omega_0,\tau_0),
\end{equation}
while $\bar{E}(N_+,N_-)$ stands for the complex conjugate of $E(N_+,N_-)$.
In particular, the lowest Landau sector energy eigenvalues are
\begin{equation}
E(N_+,N_-=0)=\hbar\frac{R_0e^{i\varphi_0}-1}{2i\tau_0}\,\left(N_++\frac{1}{2}\right)+\Delta E_0(\omega_0,\tau_0).
\end{equation}

\subsection{The $\tau_0\rightarrow 0^+$ limit}

In the absence of the interaction energy $h(\hat{\phi}^a)$, namely when $\omega_0=0$, the energy spectrum
reduces to,
\begin{equation}
\omega_0=0:\qquad E(N_+,N_-)=\frac{\hbar}{i\tau_0}\,N_-,
\end{equation}
displaying the infinite degeneracy in $N_+=0,1,2,\ldots$ of the Landau levels labelled by $N_-=0,1,2,\ldots$ and
separated by a gap $\hbar/(i\tau_0)$ as expected, then corresponding to the states $|n_+=N_+,n_-=N_-;\Omega\rangle$.
In the limit that $\tau_0=0$, only the lowest Landau level retains a finite (vanishing) energy.

When the interaction energy $h(\hat{\phi}^a)$ is included, the gap between Landau sectors is determined by the
quantity
\begin{equation}
\hbar\frac{R_0e^{i\varphi_0}+1}{2i\tau_0},
\end{equation}
which in the limit $\tau_0\rightarrow 0^+$ behaves as,
\begin{equation}
\hbar\frac{R_0e^{i\varphi_0}+1}{2i\tau_0}\stackrel{\tau_0\rightarrow 0^+}{\simeq} \frac{\hbar}{i\tau_0}+\hbar\omega_0+\cdots .
\end{equation}
Hence once again it is the scale $\hbar/(i\tau_0)$ which sets the leading contribution to that gap,
which diverges in the limit $\tau_0\rightarrow 0^+$. Consequently, only the Landau sector with $N_-=0$
retains a finite energy in that limit. Furthermore within a given Landau sector, the spacing between
states is determined by the second relevant quantity,
\begin{equation}
\hbar\frac{R_0e^{i\varphi_0}-1}{2i\tau_0},
\end{equation}
which in the limit $\tau_0\rightarrow 0^+$ behaves as,
\begin{equation}
\hbar\frac{R_0e^{i\varphi_0}-1}{2i\tau_0}\stackrel{\tau_0\rightarrow 0^+}{\simeq} \hbar\omega_0+\cdots.
\end{equation}
Hence, in that limit, the energy spectrum behaves as,
\begin{equation}
E(N_+,N_-)\stackrel{\tau_0\rightarrow 0^+}{\simeq}\frac{\hbar}{i\tau_0}\left(1+i\omega_0\tau_0+\cdots\right)\,N_-\,+\,
\left(\hbar\omega_0+\cdots\right)\left(N_+ +\frac{1}{2}\right)\,+\,\cdots.
\end{equation}
Those states retaining a finite energy in that limit belong only to the lowest Landau sector with $N_-=0$,
\begin{equation}
\lim_{\tau_0\rightarrow 0^+}E(N_+,N_-=0)=\hbar\omega_0\left(N_++\frac{1}{2}\right).
\end{equation}
In this expression one recognizes the real energy spectrum of the harmonic oscillator, including its
quantum vacuum energy, the corresponding energy eigenstates being the Fock states $|n_+=N_+,n_-=0;\Omega\rangle$.
Hence indeed the subspace of the extended Hilbert space of the extended system spanned by the lowest Landau sector
in the limit $\tau_0=0$ determines the Hilbert space of the original quantum system, in the present case that of the harmonic oscillator.

To show that the remaining Landau sectors do decouple from the energy spectrum in the limit $\tau_0=0$, it suffices
to consider the quantum evolution operator of the extended system. Given the spectral resolution of the unit operator
in terms of the eigenstates of the Hamiltonian operator, $\hat{H}$, and its adjoint, one has for the evolution
operator with $T=t_f-t_i>0$,
\begin{equation}
e^{-\frac{i}{\hbar}T\hat{H}}=\sum_{N_+,N_-=0}^\infty|N_+,N_-;\Omega_A\rangle\,e^{-\frac{i}{\hbar}TE(N_+,N_-)}\,
\langle N_+,N_-;\Omega_B|.
\end{equation}
Using the above expansion in $\tau_0$ for $E(N_+,N_-)$, one thus finds that all the states with $N_-\ge 1$ decouple exponentially in the
considered limit, 
\begin{equation}
\lim_{\tau_0\rightarrow 0^+} e^{-\frac{i}{\hbar}T\hat{H}}\stackrel{T>0}{=}
\sum_{N_+=0}^\infty|N_+,N_-=0;\Omega\rangle\,e^{-i\omega_0T(N_++1/2)}\,\langle N_+,N_-=0;\Omega|.
\end{equation}
Hence indeed all but the states belonging to the lowest Landau sector have decoupled from the dynamics of the
extended system in the limit $\tau_0\rightarrow 0^+$, leaving over precisely the Hilbert space  of the ordinary harmonic oscillator
with the correct energy spectrum and quantum time evolution operator. The states $|n_+,n_-=0;\Omega\rangle$
correspond exactly to the usual Fock states $|n_+\rangle$ of the harmonic oscillator with energy spectrum
$E(n_+)=\hbar\omega_0(n_++1/2)$, $|n_+,n_-=0;\Omega\rangle\equiv|n_+\rangle$.

This conclusion is thus in full accord with the general discussion and results of Ref.\cite{KD1.Klauder1} within the
functional integral setting, but achieved in the specific case of the harmonic oscillator and using rather
operator quantisation techniques. As a matter
of fact, a similar analysis based on the operator quantisation of the extended system for whatever initial system
and given (polynomial \cite{KD1.Klauder1})
Hamiltonian $H_0(q,p)$ is possible, leading of course to the same general conclusion\cite{KD1.Mato,KD1.GovMato}.

The above has thus also established that the limit $\tau_0\rightarrow 0^+$ enforces the projection
effected by the operator $\mathbb{P}_0$ introduced previously, thereby leading back to noncommuting hermitian
projected phase space operators, $Q=\mathbb{P}_0\hat{q}\mathbb{P}_0$ and $P=\mathbb{P}_0\hat{p}\mathbb{P}_0$,
obeying the usual Heisenberg algebra as it should, $\left[Q,P\right]=i\hbar\mathbb{P}_0$, of which the projected Hilbert space
spanned by the Fock states $|n\rangle\equiv|n,0;\Omega\rangle$ provides the usual Fock space representation. However, before the projection
is effected, as two dimensional configuration space operators, the unprojected coordinates $\hat{q}$ and $\hat{p}$ acting on the
extended Hilbert space are commuting operators. This feature, unique to the Klauder--Daubechies construction of the
phase space path integral which is covariant under canonical transformations of the original system, can be put to use
to exploit at the quantum level all the advantages of classical action-angle transformations for systems which are integrable
in the Liouville sense and which possess nonperturbative configurations \cite{KD1.Mato,KD1.GovMato}.

\section{The Deformed Harmonic Oscillator}
\label{KD1.Sec4}

\subsection{A deformed quantum dynamics}

The previous discussion has thus established that one has for the $\mathbb{P}_0$ projected evolution operator of the
extended system, when $T>0$,
\begin{equation}
\lim_{\tau_0\rightarrow 0^+}\mathbb{P}_0\,e^{-\frac{i}{\hbar}T\hat{H}}\mathbb{P}_0=
\lim_{\tau_0\rightarrow 0^+}e^{-\frac{i}{\hbar}T\hat{H}},
\end{equation}
the latter quantity then reproducing the quantum evolution operator of the original system.
However since $\mathbb{P}_0$ effects the projection onto the Hilbert space of the original quantum system,
it may be worth considering the projected evolution operator also for a finite value of $\tau_0$,
\begin{equation}
T>0:\qquad \mathbb{U}(T)=\mathbb{P}_0\,e^{-\frac{i}{\hbar}T\hat{H}}\,\mathbb{P}_0\ne
\lim_{\tau_0\rightarrow 0^+}\mathbb{P}_0\,e^{-\frac{i}{\hbar}T\hat{H}}\mathbb{P}_0,
\end{equation}
knowing that in the limit $\tau_0=0$ this operator reproduces the correct evolution operator
of the original quantum system,
\begin{equation}
U(T)=\lim_{\tau_0\rightarrow 0^+}\mathbb{U}(T)=\lim_{\tau_0\rightarrow 0^+} e^{-\frac{i}{\hbar}T\hat{H}}.
\end{equation}

Keeping $\tau_0$ finite for $\mathbb{U}(T)$ thus induces a deformed quantum dynamics inside the Hilbert space
of the original quantum system, as compared to the operator $U(T)$. Such a deformation may be of
physical interest, in a spirit comparable to that which suggests to consider noncommutative deformations
of the geometrical properties of spacetime in attempts towards formulations for a quantum theory of gravity 
through deformations of quantum algebras \cite{KD1.GovMatt,KD1.Matt,KD1.SG}.
Nevertheless, it should be pointed out that for a finite value of $\tau_0$, because of the irreversible character
of its Brownian motion component, such a dynamics is no longer unitary,
\begin{equation}
\mathbb{U}^\dagger(T)\ne \mathbb{U}^{-1}(T),\qquad
\mathbb{U}^\dagger(T)\,\mathbb{U}(T)\ne \mathbb{P}_0,\qquad
\mathbb{U}(T)\,\mathbb{U}^\dagger(T)\ne\mathbb{P}_0,
\end{equation}
and thus cannot preserve quantum probabilities, or more correctly in the present context, the total occupation number 
(the sum of the occupation densities over all quantum states of the system).
Nor does it meet the usual convolution property under consecutive time evolution intervals,
\begin{equation}
\mathbb{U}(T_2)\cdot\mathbb{U}(T_1)\ne\mathbb{U}(T_2+T_1),\qquad
\mathbb{P}_0\,e^{-\frac{i}{\hbar}T_2\hat{H}}\,\mathbb{P}_0\cdot\mathbb{P}_0\,e^{-\frac{i}{\hbar}T_1\hat{H}}\,\mathbb{P}_0\ne
\mathbb{P}_0\,e^{-\frac{i}{\hbar}(T_2+T_1)\hat{H}}\,\mathbb{P}_0.
\end{equation}
Hence such a proposal raises a series of interpretational issues, which we shall not attempt to
address here. However let us point out that when extrapolated to a quantum field theory context \cite{KD1.GovMatt2},
a finite $\tau_0$ value provides in effect a regularisation of short-distance singularities, akin to a
soft exponential cut-off in the momentum of quantum states, indeed so efficient that all quantum amplitudes for whatever
field theory in a perturbative expansion, even including general relativity, are ultra-violet finite (the only potential source
of trouble being some tadpole contributions, which may always be dealt with by a proper choice of quantum Hamiltonian). The combination
of the time scale $\tau_0$---expected to be extremely small as well if non vanishing  in the physical world---and of
the Planck time in a quantum gravitational context, $\tau_{\rm Planck}=\sqrt{\hbar G_N/c^5}\simeq 10^{-43}$~s---irrespective
of whether these two time scales should prove to be unrelated or not---, may thus offer some tantalising prospects for strongly gravitationally
interacting quantum systems \cite{KD1.GovMatt}, a physical situation in which perhaps the requirements of unitarity
and Lorentz invariance may be relaxed to some slight degree for what concerns
experimentally unexplored extreme regimes. Whatever the case may be, at least a nonvanishing time scale $\tau_0$
provides yet another regularisation of short-distance quantum dynamics for local field theories whose usefulness
is worth exploring.

As a matter of fact the projected operator $\mathbb{U}(T)$ has already been computed \cite{KD1.GovMatt,KD1.Matt} directly from
the KD-PI in (\ref{eq:KDPI1}) using a saddle point approach for what is indeed a purely gaussian functional integral in the case of the
harmonic oscillator. Here rather, we shall exploit the operator solution constructed above to reproduce the
same result, making it readily explicit that the deformed quantum dynamics remains diagonal in the Fock state basis
of the harmonic oscillator.

\subsection{The projected evolution operator}

Since the operator of interest is of the form
\begin{equation}
\mathbb{U}(T)\stackrel{T>0}{=}\sum_{n_+,m_+=0}^\infty |n_+,0;\Omega\rangle\langle n_+,0;\Omega| e^{-\frac{i}{\hbar}T\hat{H}}
|m_+,0;\Omega\rangle\langle m_+,0;\Omega|,
\end{equation}
while the eigenstates of $\hat{H}$ (resp., $\hat{H}^\dagger$) are $|N_+,N_-;\Omega_A\rangle$
(resp., $|N_+,N_-;\Omega_B\rangle$), one first needs to consider the following change of basis matrix elements,
\begin{equation}
\langle n_+,0;\Omega|N_+,N_-;\Omega_A\rangle,\qquad
\langle m_+,0;\Omega|N_+,N_-;\Omega_B\rangle.
\end{equation}

Using the definition of the $A$-Fock states $|N_+,N_-;\Omega_A\rangle$ and the representation of $|\Omega_A\rangle$
as a coherent helicity pairing excitation of $|\Omega\rangle$, a detailed evaluation of the first matrix element finds the
following result,
\begin{equation}
\langle n_+,0;\Omega|N_+,N_-;\Omega_A\rangle=
\left(\frac{2}{\rho+\rho^{-1}}\right)\left(\frac{2}{\rho+\rho^{-1}}\right)^{N_+}
\left(\frac{\rho-\rho^{-1}}{2}\right)^{N_-}\sqrt{\frac{N_+!}{n_+!\, N_-!}}\,\delta_{N_+,N_-+n_+}.
\end{equation}
In a similar fashion,
\begin{equation}
\langle m_+,0;\Omega|N_+,N_-;\Omega_B\rangle=
\left(\frac{2}{\bar{\rho}+\bar{\rho}^{-1}}\right)
\left(\frac{2}{\bar{\rho}+\bar{\rho}^{-1}}\right)^{N_+}
\left(\frac{\bar{\rho}-\bar{\rho}^{-1}}{2}\right)^{N_-}
\sqrt{\frac{N_+!}{m_+!\, N_-!}}\,\delta_{N_+,N_-+m_+}.
\end{equation}

It then readily follows that the matrix elements of the deformed evolution operator $\mathbb{U}(T)$ in
the Fock state basis of the harmonic oscillator are diagonal in that basis,
\begin{equation}
\langle n|\mathbb{U}(T)|\ell\rangle\equiv
\langle n,0;\Omega|\mathbb{U}(T)|\ell,0;\Omega\rangle=
\delta_{n,\ell}\langle n|\mathbb{U}(T)|n\rangle.
\end{equation}
Using the above results, a direct evaluation of the diagonal matrix element then leads to,
\begin{equation}
\langle n|\mathbb{U}(T)|n\rangle=
e^{-\frac{i}{\hbar}T\Delta E_0}\,e^{-i(n+\frac{1}{2})\alpha_+}\,F^{n+1}(T),
\end{equation}
where,
\begin{equation}
\alpha_+=\omega_0T\,\frac{R_0e^{i\varphi_0}-1}{2i\omega_0\tau_0},\qquad
\alpha_-=\omega_0T\,\frac{R_0e^{i\varphi_0}+1}{2i\omega_0\tau_0},
\end{equation}
and,
\begin{equation}
\frac{1}{F(T)}=\left(\frac{\rho+\rho^{-1}}{2}\right)^2-\left(\frac{\rho-\rho^{-1}}{2}\right)^2 e^{-i(\alpha_+ + \alpha_-)}.
\end{equation}

In order to bring this matrix element to a more amenable form, in terms of the two quantities $R_0$ and $\varphi_0$
defined previously already through the identification (\ref{eq:R0P0}) let us introduce the following further notations,
\begin{equation}
R=\sqrt{\frac{1}{2}\left(R^2_0+1\right)},\qquad
S=\frac{1}{2}\left(R+1\right),
\end{equation}
which are such that
\begin{equation}
R-1=2\frac{\omega^2_0\tau^2_0}{R^2 S},\qquad
\frac{\omega_0\tau_0}{RS}=\sqrt{1-\frac{1}{S}},\qquad
\cos\varphi_0=\frac{R}{R_0},\qquad
\sin\varphi_0=\frac{2\omega_0\tau_0}{R_0 R}.
\end{equation}
It then follows that,
\begin{equation}
i\alpha_+=T\frac{R-1}{2\tau_0}+i\frac{\omega_0 T}{R},\qquad
i\left(\alpha_+ + \alpha_-\right)=T\frac{R}{\tau_0}+2i\frac{\omega_0T}{R},
\end{equation}
as well as,
\begin{equation}
\rho^2=R+2i\frac{\omega_0\tau_0}{R},\qquad
\rho^{-2}=\frac{R^2-2i\omega_0\tau_0}{R^2_0 R},
\end{equation}
and finally,
\begin{equation}
\frac{1}{F(T)}=e^{-\frac{R}{\tau_0}T}\,e^{-2i\frac{\omega_0}{R}T}\,+\,
S\frac{R+2i\omega_0\tau_0}{R^2+2i\omega_0\tau_0}
\left(1-e^{-\frac{R}{\tau_0}T}\,e^{-2i\frac{\omega_0}{R}T}\right).
\end{equation}
Hence we have so far,
\begin{equation}
\langle n|\mathbb{U}(T)|n\rangle=e^{-\frac{i}{\hbar}T\Delta E_0(\omega_0,\tau_0)}\,
e^{-i\frac{\omega_0 T}{R}\left(n+\frac{1}{2}\right)}\,
e^{-\frac{R-1}{2\tau_0}T\left(n+\frac{1}{2}\right)}\,F^{n+1}(T).
\end{equation}

Since
\begin{equation}
\lim_{T\rightarrow +\infty}F(T)=\frac{1}{S}\frac{R^2+2i\omega_0\tau_0}{R+2i\omega_0\tau_0},
\end{equation}
in order that the asymptotic time limit $T\rightarrow +\infty$ leaves over at least one of the matrix
elements $\langle n|\mathbb{U}(T)|n\rangle$ with a finite and non vanishing
occupation, given that $R>1$ this can only be the case for the Fock vacuum
$|n=0\rangle=|\Omega\rangle$, which requires then to specify the choice for the arbitrary function $\Delta E_0(\omega_0,\tau_0)$ as follows,
\begin{equation}
\Delta E_0(\omega_0,\tau_0)=-i\hbar\frac{R-1}{4\tau_0}\stackrel{\tau_0\rightarrow 0^+}{\simeq}
-\frac{1}{2}i\hbar\omega^2_0\tau_0+\ldots,
\end{equation}
indeed a pure imaginary quantity but such that it vanishes in the limit $\tau_0=0$, as it should.
Correspondingly, we have for the energy subtraction constant $E_0$,
\begin{equation}
E_0=\hbar\frac{R_0e^{i\varphi_0}-R+2}{4i\tau_0}=\hbar\frac{1}{2i\tau_0}+\hbar\frac{\omega_0}{2R}.
\end{equation}
Incidentally, the exact same choice had to be made in Refs.\cite{KD1.GovMatt,KD1.Matt} for precisely the same reason.
Note also that in the absence of the interaction energy $h(\hat{\phi}^a)$,
namely when $\omega_0=0$, the value $E_0=\hbar/(2i\tau_0)$ coincides precisely with the one specified in Ref.\cite{KD1.Klauder1}
for the factor $e^{C_0T/\tau_0}$ in (\ref{eq:KDPI1}).

In conclusion, the final expression for the relevant matrix elements, which agrees with the result obtained through
a functional integral calculation \cite{KD1.GovMatt,KD1.Matt}, is,
\begin{equation}
\langle n|\mathbb{U}(T)|\ell\rangle=\delta_{n,\ell}\cdot
e^{-i\frac{\omega_0}{R}T\left(n+\frac{1}{2}\right)}\,
e^{-n\frac{R-1}{2\tau_0}T}\,F^{n+1}(T)\equiv \delta_{n,\ell}\cdot\mathbb{U}_n(T),
\end{equation}
hence,
\begin{equation}
\mathbb{U}(T)=\sum_{n=0}^\infty |n\rangle\,e^{-i\frac{\omega_0}{R}T\left(n+\frac{1}{2}\right)}\,
e^{-n\frac{R-1}{2\tau_0}T}\,F^{n+1}(T)\,\langle n|=\sum_{n=0}^\infty |n\rangle\,\mathbb{U}_n(T)\,\langle n|.
\end{equation}
Note that we have
\begin{equation}
\lim_{T\rightarrow 0^+}\mathbb{U}(T)=\sum_{n=0}^\infty |n\rangle\langle n|=\mathbb{P}_0,\qquad
\lim_{\tau_0\rightarrow 0^+}\mathbb{U}(T)=\sum_{n=0}^\infty |n\rangle\,e^{-i\omega_0 T(n+\frac{1}{2})}\,\langle n|
=U(T),
\end{equation}
as it should, while,
\begin{equation}
\lim_{T\rightarrow +\infty}\mathbb{U}(T)=e^{-\frac{1}{2}i\frac{\omega_0}{R}T}\,
\frac{1}{S}\frac{R^2+2i\omega_0\tau_0}{R+2i\omega_0\tau_0}\,|\Omega\rangle\langle\Omega|,
\label{eq:decayFock}
\end{equation}
thus displaying how because of the Brownian motion contribution to the quantum dynamics when $\tau_0\ne 0$,
whatever the initial state of the system it eventually decays to the Fock vacuum with a specific factor rescaling
the initial occupation of that particular state.

Before commenting on the significance of these results, let us consider how the original Heisenberg algebra
of phase space operators is deformed in the time evolved picture of the system, because of a non vanishing
value for $\tau_0>0$. Defining quantum operators, $A(t_f)$, in the Heisenberg picture in the usual way but in terms
of the projected evolution operator, $\mathbb{U}(T)$ with $T=t_f-t_i>0$, as,
\begin{equation}
A(t_f)=\mathbb{U}^\dagger(T)\,A(t_i)\,\mathbb{U}(T),
\end{equation}
a direct calculation in the case of the (projected) position and momentum operators, $Q(t_i)=\mathbb{P}_0\hat{q}(t_i)\mathbb{P}_0$
and $P(t_i)=\mathbb{P}_0\hat{p}(t_i)\mathbb{P}_0$
with $\left[Q(t_i),P(t_i)\right]=i\hbar\mathbb{P}_0$, finds indeed a deformed Heisenberg algebra,
\begin{eqnarray}
\left[Q(t_f),P(t_f)\right] &=&\quad i\hbar\,|0\rangle\, F^3_0\,D^3(T)\,e^{-\frac{R-1}{\tau_0}T}\,\langle 0|\,+  \\
&& +\,i\hbar\sum_{n=1}^\infty\,|n\rangle\,F^{2n+1}_0\,D^{2n+1}(T)\,e^{-(2n-1)\frac{R-1}{\tau_0}T}\,
\left[\left(n+1\right)F^2_0\,D^2(T)\,e^{-2\frac{R-1}{\tau_0}T}\,-\,n\right]\,\langle n|. \nonumber
\end{eqnarray}
In this expression the quantities $F_0$ and $D(T)$ are defined according to the relation
\begin{equation}
|F(T)|^2=F_0\cdot D(T),
\end{equation}
where
\begin{equation}
F_0=\frac{1}{S^2}\,\frac{R^4+4\omega^2_0\tau^2_0}{R^2+4\omega^2_0\tau^2_0},
\label{eq:F0}
\end{equation}
and,
\begin{equation}
\frac{1}{D(T)}=1+2\left(\frac{R-1}{R+1}\right)\,e^{-\frac{R}{\tau_0}T}\,\cos 2\left(\frac{\omega_0}{R}T+\varphi_0\right)\,+\,
\left(\frac{R-1}{R+1}\right)^2\,e^{-\frac{2R}{\tau_0}T}.
\label{eq:DT}
\end{equation}
Note that we have,
\begin{equation}
\lim_{T\rightarrow 0^+}\left[Q(t_f),P(t_f)\right]=i\hbar\mathbb{P}_0,\qquad
\lim_{\tau_0\rightarrow 0^+}\left[Q(t_f),P(t_f)\right]=i\hbar\mathbb{P}_0,
\end{equation}
as it should, while,
\begin{equation}
\lim_{T\rightarrow +\infty}\left[Q(t_f),P(t_f)\right]=0.
\end{equation}

The reason why in the asymptotic time limit, $T\rightarrow +\infty$, the two phase space operators $Q(t_f)$ and $P(t_f)$
end up commuting with one another as in the classical system, is that in that limit all quantum states of the harmonic
oscillator except for its Fock vacuum have exponentially decayed to zero, as shown explicitly by (\ref{eq:decayFock}).

\subsection{Physical implications}

More precisely, given an initial quantum state
\begin{equation}
|\psi,t_i\rangle=\sum_{n=0}^\infty|n\rangle\,\psi_n(t_i),\qquad \psi_n(t_i)\in\mathbb{C},\qquad
\sum_{n=0}^\infty|\psi_n(t_i)|^2<\infty,
\end{equation}
its configuration at time $t_f$ with $T=t_f-t_i>0$ is
\begin{equation}
|\psi,t_f\rangle=\mathbb{U}(T)|\psi,t_i\rangle=\sum_{n=0}^\infty |n\rangle\,\mathbb{U}_n(T)\,\psi_n(t_i)=
\sum_{n=0}^\infty |n\rangle\,\psi_n(t_f),\qquad
\psi_n(t_f)=\mathbb{U}_n(t_f-t_i)\,\psi_n(t_i).
\end{equation}
Consequently, the time evolution of the occupation densities of the Fock eigenstates of the harmonic
oscillator is determined by,
\begin{equation}
|\psi_n(t_f)|^2=|\mathbb{U}_n(t_f-t_i)|^2\cdot|\psi_n(t_i)|^2.
\end{equation}
Based on the expressions above, one has
\begin{equation}
|\mathbb{U}_n(T)|^2=|F(T)|^{2(n+1)}\,e^{-n\frac{R-1}{\tau_0}T},
\end{equation}
namely,
\begin{equation}
|\mathbb{U}_n(T)|^2=F^{n+1}_0\,D^{n+1}(T)\,e^{-n\frac{R-1}{\tau_0}T},
\end{equation}
where the quantities $F_0$ and $D(T)$ are given in (\ref{eq:F0}) and (\ref{eq:DT}), respectively.

Stochastic Brownian motion leads to so efficient a statistical decoherence of the quantum system that whatever dynamics
there is to begin with, it totally decays away. All that remains in a rescaled occupation of the initial ground state occupation of the system.
Given the asymptotic values,
\begin{equation}
\lim_{T\rightarrow +\infty}\left|\mathbb{U}_{n=0}(T)\right|^2=\frac{1}{S^2}\frac{R^4+4\omega^2_0\tau^2_0}{R^2+4\omega^2_0\tau^2_0},\qquad
\lim_{T\rightarrow +\infty}\left|\mathbb{U}_{n\ge 1}(T)\right|^2=0,
\end{equation}
the time asymptotics of the Fock state occupations is such that,
\begin{equation}
\lim_{t_f\rightarrow +\infty}|\psi_{n=0}(t_f)|^2=
\frac{1}{S^2}\frac{R^4+4\omega^2_0\tau^2_0}{R^2+4\omega^2_0\tau^2_0}\ |\psi_{n=0}(t_i)|^2,\qquad
\lim_{t_f\rightarrow +\infty}|\psi_{n\ge 1}(t_f)|^2=0.
\end{equation}
In its large time behaviour, the dynamics of the (non interacting closed) system which is irreversible provided $\tau_0$ is non vanishing
however small its value, is such that the Hilbert space of the quantum system thus becomes effectively one dimensional, being
aligned along the direction only of the oscillator Fock vacuum $|\Omega\rangle=|0\rangle$. All other excited Fock states
$|n\rangle$ decouple by decay (without being coupled to some external environment or interaction)
with a hierarchy of lifetimes determined by $\tau^{(n)}_+=\tau_0/(n(R-1))$, $n=1,2,\ldots$

More specifically, first one observes an oscillatory pattern contributing both to the overall phase factor proportional to $(n+1/2)$
in $\mathbb{U}_n(T)$ and to the function $F(T)$, and thus to its modulus squared $|F(T)|^2=F_0 D(T)$ in $|\mathbb{U}_n(T)|^2$.
The periodicity of this pattern is set by a rescaling of the proper time scale of the oscillator by the factor $R$,
namely by the following effective angular frequency,
\begin{equation}
\omega_{\rm effective}=\frac{\omega_0}{R}<\omega_0.
\end{equation}
Besides this oscillatory pattern, the time dependence of the Fock state occupations, modulated by $\left|\mathbb{U}_n(T)\right|^2$, is
furthermore governed by two more real exponential time scales, the first of which modulates the factor $|F(T)|^2$ and the second which modulates
the exponential in time normalisation of $\left|\mathbb{U}_n(T)\right|^2$ for $n\ge 1$,
\begin{equation}
\tau_-=\frac{\tau_0}{R}\stackrel{\tau_0\rightarrow 0^+}{\simeq}\tau_0+\cdots,\qquad
\tau^{(n)}_+=\frac{1}{n}\frac{\tau_0}{R-1}=\frac{1}{n}\frac{R^2S}{2\omega^2_0\tau_0}\stackrel{\tau_0\rightarrow 0^+}{\simeq}
\frac{1}{n}\frac{1}{2\omega^2_0\tau_0}+\cdots
\end{equation}
or, when measured in units either of the characteristic time scale of the oscillator, $1/\omega_0$, or the intrinsic time
scale $\tau_0$ of the quantum deformation of its dynamics,
\begin{equation}
\begin{array}{rclcl}
\omega_0\tau_- &=& \frac{\omega_0\tau_0}{R}\stackrel{\tau_0\rightarrow 0^+}{\simeq}\omega_0\tau_0+\cdots &,&
\omega_0\tau^{(n)}_+=\frac{1}{n}\frac{\omega_0\tau_0}{R-1}=\frac{1}{n}\frac{R^2 S}{2\omega_0\tau_0}\stackrel{\tau_0\rightarrow 0^+}{\simeq}
\frac{1}{n}\frac{1}{2\omega_0\tau_0}+\cdots, \\
 && && \\
\frac{\tau_-}{\tau_0} &=& \frac{1}{R}\stackrel{\tau_0\rightarrow 0^+}{\simeq}1+\cdots &,&
\frac{\tau^{(n)}_+}{\tau_0}=\frac{1}{n}\frac{1}{R-1}=\frac{1}{n}\frac{R^2S}{2(\omega_0\tau_0)^2}\stackrel{\tau_0\rightarrow 0^+}{\simeq}
\frac{1}{n}\frac{1}{2(\omega_0\tau_0)^2}+\cdots .
\end{array}
\end{equation}

For a given value of $\omega_0\tau_0$, and provided $n$ is small enough such that $\tau_-<\tau^{(n)}_+$ (which is
always the case for $n=1$ at least), for any given Fock state $|n\rangle$ there are then
effectively three time windows characteristic of different regimes for the deformed quantum dynamics, namely
$0\le T\le\tau_-$, $\tau_-\le T\le\tau^{(n)}_+$ and $\tau^{(n)}_+\le T<\infty$ \cite{KD1.GovMatt,KD1.Matt}. To describe these windows
it is relevant to consider the value of the characteristic time scale of the system, $1/\omega_0$, relative
to the time scale of the deformation, namely the quantity $1/(\omega_0\tau_0)$ (note that if the physical
system under consideration does not carry any characteristic time scale, for instance a free particle, no deviation
from ordinary unitary quantum dynamics is present even when $\tau_0\ne 0$). Since ordinary quantum behaviour
is recovered in the limit $\tau_0\rightarrow 0^+$, when the quantity $1/(\omega_0\tau_0)$ is extremely large, for
all practical purposes the quantum behaviour of the system does not significantly differ from that of ordinary
quantum mechanics, at least up to the time scale $\tau^{(n)}_+$ for each of those Fock states $|n\rangle$ such that
$\tau^{(n)}_+>\tau_-$. In the time window $\tau_-\le T\le\tau^{(n)}_+$, only a very small time dependent rescaling of the
Fock state occupation is occurring which is the less perceptible the larger is the value of $1/(\omega_0\tau_0)$.
Since if indeed non vanishing in the physical world the actual value of $\tau_0$ is expected to be on the order of the Planck time,
some $10^{-43}$~s, while in comparison experimental conditions have not yet observed 
extremely high intensity excitations of modes of large enough frequencies for particle and interaction fields, it seems fair to
assume that until now all experiments conducted in laboratories have remained inside this ``ordinary quantum physics window"
(this does not include violent astrophysical phenomena in strong gravitational quantum regimes that may be observed).
It is only by moving into time scales $1/\omega_0$ becoming comparable to $\tau_0$, that the time window for
ordinary quantum mechanics begins to grow narrow enough that the deformed quantum dynamics of the system may
start display deviations from ordinary unitary
quantum behaviour, and thereby enable at least experimental upper bounds to be set on the deformation parameter $\tau_0$.

When reaching such a regime, which is then essentially also the situation for those Fock states $|n\rangle$ with $n$
sufficiently large such that now $\tau^{(n)}_+<\tau_-$, as well as for the time window $\tau^{(n)}_+\le T<\infty$
even in the discussion above, the telltale signs for the lack of a unitary quantum dynamics are,
first, the total decoherence of the dynamics decaying ultimately to its ground state (on a time scale which is the
smaller the larger is $\omega_0\tau_0$), and second, the time dependent rescaling or renormalisation of the occupation density
of that ground state and of the excited states at intermediate times, with in particular for the ground state an asymptotic in
time rescaling of its occupation given by the quantity
\begin{equation}
F_0=\frac{1}{S^2}\,\frac{R^4+4\omega^2_0\tau^2_0}{R^2+4\omega^2_0\tau^2_0}.
\end{equation}
The behaviour of the latter factor as a function of $1/(\omega_0\tau_0)$ is noteworthy\cite{KD1.GovMatt,KD1.Matt},
\begin{equation}
F_0\stackrel{1/(\omega_0\tau_0)\rightarrow 0}{\longrightarrow} 4\,\frac{1}{\omega_0\tau_0}+\ldots,\qquad
F_0\stackrel{1/(\omega_0\tau_0)\rightarrow +\infty}{\longrightarrow} 1+\frac{2}{\left(1/(\omega_0\tau_0)\right)^2}+\ldots
\end{equation}
Hence, as $\tau_0\rightarrow 0^+$, the population rescaling factor $F_0$ keeps on approaching the unit value
it has when $\tau_0=0$ {\it but from above\/}, which means that as $1/(\omega_0\tau_0)$ decreases $F_0$ keeps
on growing ever larger than unity, until it reaches a maximal value lying above unity ($F^{\rm max}_0\simeq 1.079$ for
$1/(\omega_0\tau_0)\simeq 2.591$) and from which further on,
as $1/(\omega_0\tau_0)$ still keeps decreasing, $F_0$ starts decreasing as well, then passes the unit value,
to finally reach a vanishing value in the limit that $1/(\omega_0\tau_0)$ also vanishes.
Consequently given a value for $\tau_0$, for an angular frequency larger than a certain threshold, $\omega_{\rm threshold}(\tau_0)$,
the survival occupation density of even the Fock vacuum is always less than its initial
value, while for $\omega_0$ values less than $\omega_{\rm threshold}(\tau_0)$, the survival occupation density is always
larger than its initial value. Nonetheless in all circumstances all excited Fock states end up not being populated
at all at asymptotic times. Within a quantum field theory context, especially for the gravitational field, clearly such behaviour
implies some tantalising prospects for dynamics at the smallest spacetime scales, leading to an effective
coarse-graining of spacetime geometry since this geometry may only be probed through interacting quantum fields.

\section{Conclusions}
\label{KD1.Sec5}

This paper considered the canonical operator quantisation formalism corresponding to the functional
integral of the Klauder--Daubechies construction of the phase space path integral\cite{KD1.Klauder1}.
The latter formulation introduces a regularisation parameter, equivalent to a new time scale $\tau_0>0$, such that in the limit where
it vanishes the construction reproduces the correct quantum dynamics of the system. This result was
demonstrated explicitly from the operator representation of the same construction, in the specific case
of the harmonic oscillator, thereby highlithing from a different and complementary point of view the
inner workings of the Klauder--Daubechies approach to quantum dynamics.

In effect, this approach promotes the original system to the dynamics of an extended one of which the
configuration space is the phase space of the original system, equipped not only with that phase space's
symplectic geometry but also a Riemannian metric with identical volume form. The latter structure is related
to a Brownian motion component added to the quantum dynamics of the original system, such that when
the Brownian motion regularisation is taken away again, only the original quantum system survives.
This formulation offers a number of advantages, not least of which is its manifest covariance under
general canonical transformations of the phase space parametrisation, which may be put to efficient
use to develop new nonpertubative quantisation techniques \cite{KD1.GovMato,KD1.Mato}. Furthermore the
extended regularising dynamics is of the form of a generalised Landau problem in phase space,
with a pure positive imaginary mass set by the time scale parameter $\tau_0$. In this respect,
the Klauder--Daubechies construction comes in close resonance with present day developments
in noncommutative geometry and quantum mechanics, most of which are inspired precisely by the
Landau problem in the plane in which the mass parameter is taken to vanish \cite{KD1.SG}.

The operator formulation of the Klauder--Daubechies construction should
also make it possible to extend it to systems with more than a single degree of freedom, one first
case of interest being precisely the Landau problem itself and its associated noncommutative geometry of
the Moyal plane. But beyond that, relativistic quantum field theories with
their short-distance divergences in perturbation theory are another case in point. Indeed, the operator
technique is well adapted to keep the value of $\tau_0$ finite throughout, which is possibly a choice of physical relevance
in the spirit of deformations of quantum algebraic structures, which however then reveals some appealing as well as
some not so appealing new features. If only for that purpose, a finite $\tau_0$ provides a new type of short-distance
regularisation in local quantum field theory taming all short-distance divergences. On the other hand, unitarity and
Lorentz invariance are then lost at time scales less than $\tau_0$, with however a suppression of dynamics
precisely on those scales as well which is bound to induce an effective coarse-graining of spacetime geometry
in strong gravitational quantum systems. In the latter context, the status of initial cosmological singularities, or
the issue of trans-Planckian energies in black hole radiation are open issues that come to mind, which could be
addressed within the Klauder--Daubechies framework for quantum dynamics.

\section*{Acknowledgements}

The authors wish to address their warm words of appreciation to Prof. John R. Klauder for insightful discussions
and his constant interest in the present work. Prof. Gerhard C. Hegerfeldt is also thanked for a constructive question having
led to further clarification in the analysis of this paper.

The first part of this work was initiated in February-March 2008 while two of us were visiting the African
Institute for Mathematical Sciences (AIMS, Muizenberg, South Africa), J.G. as invited lecturer and C.M.B. as
the beneficiary of a two months Victor Rothschild Fellowship. We wish to thank Prof.~Fritz Hahne for his interest
and constant encouragements, and AIMS for its wonderful hospitality and the financial support which made our joint stay there possible.
Laure Gouba, postdoctoral Fellow at AIMS at that time, also took part in the initial stages of the analysis,
for which we acknowledge her collaboration.
Over the three years of his PhD work, C.M.B. benefited from a PhD Fellowship at the University of Kinshasa from the ``Coop\'eration
Universitaire au D\'eveloppement (CUD)" of the Universities of the French speaking Community of Belgium, which also
made three visits of three months each at the Catholic University of Louvain (Belgium) possible during that period.
C.M.B. is grateful to the CUD for this most essential support.
J.G. acknowledges the Abdus Salam International Centre for Theoretical Physics (ICTP, Trieste, Italy)
Visiting Scholar Programme in support of a Visiting Professorship at the ICMPA-UNESCO (Republic of Benin).
O.M. acknowledges partial support by the Marie Curie programme RTN MRTN-CT-2006-035505 through the University of Roma Tre (Rome, Italy).
The work of J.G. and O.M. is supported in part by the Institut Interuniversitaire des Sciences Nucl\'eaires (I.I.S.N., Belgium),
and by the Belgian Federal Office for Scientific, Technical and Cultural Affairs through
the Interuniversity Attraction Poles (IAP) P6/11.

\end{document}